\documentclass[10pt,conference,letter]{IEEEtran}
\usepackage{graphicx,color,epsfig,rotating}
\usepackage{epsf}
\usepackage{latexsym}
\usepackage{subfigure}
\usepackage{cite}
\usepackage{amssymb,amsmath}

\long\def\comment#1{}

\newfont{\bbb}{msbm10 scaled 700}

\newfont{\bb}{msbm10 scaled 1100}

\newcommand{\EE}{\mbox{\bb E}}

\newcommand{\hv}{{\bf h}}

\newcommand{\xv}{{\bf x}}

\newcommand{\Cc}{{\cal C}}

\newcommand{\Nc}{{\cal N}}

\renewcommand{\det}{{\hbox{det}}}

\newcommand{\herm}{{\sf H}}

\IEEEoverridecommandlockouts

\newcommand{\bfx}[1]{\mathbf{#1}}

\DeclareMathOperator*{\argmax}{arg\,max}

\begin{document}
\title{Multi-User MIMO with outdated CSI: \\ Training, Feedback and
  Scheduling} \author{\authorblockN{Ansuman Adhikary\authorrefmark{2}
    and Haralabos C. Papadopoulos\authorrefmark{1} and Sean
    A.~Ramprashad\authorrefmark{1} and Giuseppe
    Caire\authorrefmark{2}} \authorblockA{\authorrefmark{1} Docomo
    Innovations Inc., Palo Alto, CA 94304}
  \authorblockA{\authorrefmark{2} EE Dept., University of Southern
    California, Los Angeles, CA 90089}
\vspace{-0.25in}}

\maketitle

\begin{abstract}
Conventional MU-MIMO techniques, e.g. Linear
Zero-Forced Beamforming (LZFB), require sufficiently accurate
channel state information at the transmitter (CSIT) in order to
realize spectral efficient transmission (degree of freedom gains).
In practical settings, however, CSIT accuracy can be limited
by a number of issues including CSI estimation, CSI feedback delay between user terminals and base stations, and
the time/frequency coherence of the channel. The latter aspects
of CSIT-feedback delay and channel-dynamics can lead to significant
challenges in the deployment of efficient MU-MIMO systems.

Recently it has been shown by Maddah-Ali and Tse 
that degree of freedom gains can be realized by MU-MIMO even
when the knowledge of CSIT is completely outdated. Specifically,
outdated CSIT, albeit perfect CSIT,  is known for transmissions only after they have taken place. This aspect of insensitivity to CSIT-feedback delay is of particular interest since it allows one to reconsider MU-MIMO design in dynamic channel conditions.  
Indeed, as we show, with appropriate scheduling, and even in the context of CSI estimation and feedback errors, the proposed schemes based on outdated CSIT can have performance advantages over conventional MU MIMO in such scenarios.



\end{abstract}

\section{Introduction}
\label{intro}
We  consider a multiple-input multiple-output (MIMO) Gaussian broadcast
channel modeling the downlink of a cellular system involving a base
station (BS) with $M$ antennas and $L$ single-antenna user terminals
(UT). A channel use of such a channel is described by
\begin{equation}
y_k = \mathbf{h}_k^\herm \mathbf{x} + v_k , \ \ k = 1,\ldots,L
\end{equation}
where $y_k$ is the channel output at UT $k$, $v_k \sim
\mathcal{CN}(0,N_0)$ is white Gaussian
noise (WGN), $\mathbf{h}_k \in \mathbb{C}^{M\times 1}$ is the vector of
channel coefficients from the antenna of $k$-th UT to the BS antenna
array, and $\mathbf{x}$ is the vector of channel input symbols
transmitted by the BS. The channel input is subject to the average
power constraint $\EE[||\mathbf{x}||^2] \leq P$.

We assume that the collection of all channel vectors $\mathbf{H} =
[\mathbf{h}_1,\ldots,\mathbf{h}_L] \in \mathbb{C}^{M \times L}$,
varies in time according to a block fading model, where $\mathbf{H}$
is constant over a slot of length $T$ channel uses (which we refer to
as a coherence block), and evolves from slot to slot according to
an ergodic stationary spatially white jointly Gaussian process,
where the entries of $\mathbf{H}$ are Gaussian i.i.d. with elements
$\sim \mathcal{CN}(0,1)$.

If the CSI matrix $\mathbf{H}$ is perfectly and instantaneously known
to the transmitter (CSIT) and the receivers (CSIR), the capacity
region of the channel is obtained by MMSE-DFE beamforming and Gaussian
dirty paper coding (DPC) \cite{Caire-Shamai-TIT03,
  Viswanath-Tse-TIT2003,Vishwanath-Jindal-Goldsmith-TIT03,Yu-Cioffi-TIT04,Weingarten-Steinberg-Shamai-TIT04}.
In practice, however, both CSIR and CSIT are not known perfectly.  For example in frequency division duplex (FDD) systems,
UTs estimate CSI based on downlink pilots which are received with additive noise, and the transmitter is provided with
imperfect CSI via limited and delayed feedback from the UTs.
Given the sensitivity of DPC to CSIT
accuracy,  it follows that schemes which are more robust to CSIT accuracy such as
LZFB are the ones considered for actual deployments \cite{Yoo-Goldsmith-JSAC06}.

LZFB uses linear
precoding to serve $K$ out of $L$ users simultaneously (with $K\le
M$), and achieves for $K=M$ the maximum possible
degrees-of-freedom (DoFs) of $M$.  Furthermore, it has been shown that even
in the presence of estimation errors, if CSIT feedback is obtained in
the same coherence block and the precoder is properly designed, the
DoFs are still preserved, although there is constant gap from the
achievable rates under perfect CSIT \cite{Caire-Jindal-Kobayashi-Ravindran-TIT10}.

In general, however, due to inherent feedback delay, the channels
at the time of downlink pilot training differ from the channels at the
time of actual data transmission.
If such
channels are correlated
with a correlation coefficient of magnitude less than one, the DoFs promised by
the LZFB scheme are lost and
achievable rates saturate with increasing SNR  \cite{Caire-Jindal-Kobayashi-Ravindran-TIT10}.
Inherent changes in channels over time, and practical limits on feedback delays in some systems, therefore create practical challenges even with CSIT robust schemes.

Maddah-Ali and Tse \cite{maddah2010completely} have shown that even if the
CSIT is completely outdated (i.e., the BS has perfect knowledge of
past channels but no knowledge of the current channels), it is
possible to acquire DoFs greater than 1 by means of transmission
schemes that code across multiple quasistatic blocks. In particular,
the Maddah-Ali and Tse (MAT) scheme \cite{maddah2010completely} makes
use of multi-round transmissions and applies the techniques of
interference alignment (IA) to realize DoF gains. For example, with
$M=2$ antennas at the BS serving $K=2$ single antenna users, a DoF of
$\frac{4}{3}$ is achievable. In general the DoFs of such schemes scale as $\frac{M}{\log_e M}$, where
$M=K$ is number of transmit antennas and simultaneously
served users.

In this paper we consider several practical aspects that arise in
considering multi-round MU-MIMO schemes with outdated CSI. As a
prelude, in Sec.~\ref{sys-model} we present the system model of
interest in this paper, along with a brief description of the MAT
schemes from \cite{maddah2010completely}. In Sec.~\ref{csi-est-fb} we
study the effects of downlink training and CSI feedback on the
achievable rates. For simplicity we focus on the two-user MAT scheme
and show that, unlike conventional MU-MIMO, the achievable rates of
these MU-MIMO schemes do not saturate with outdated CSI.


In Sec.~\ref{scheduling} we develop methods for improving upon the
achievable rates provided by outdated-CSI schemes by means of
scheduling algorithms.  With $M$ transmit antennas, the DoFs are
maximized via a $K$-user MAT scheme using $K$ rounds with
$K=M$. However, performing IA at every round gives rise to noise enhancement.
Thus, with increasing number of rounds a higher signal to noise
ratio (SNR) is required for DoF gains to materialize.    As
we show, by leveraging gains obtained through scheduling,
multi-round MU-MIMO schemes, e.g., 3-round 3-user
schemes, can be made operationally attractive even at lower SNR.
To obtain such scheduling benefits requires the use of novel
packet-centric IA MU-MIMO schemes, which exploit the same principles
as the MAT scheme, but provide significantly more flexibility in
scheduling.   Simulation examples are provided
in Section \ref{simulation},
and finally, a summary with conclusions are given in Section
\ref{conclusion}.


\section{System Model and Introduction to MAT} \label{sys-model}
Throughout, we assume the presence of a sequence of quasistatic
channels between the $n$-\/th user and the transmitter.  We define
$\hv_n[t]$ as the $1\times M$ channel between the transmitter and
$n$-\/th user over the $t$-\/th quasistatic interval, which hereby is
referred to as the $t$-\/th slot. We assume that the user channel
coefficients corresponding to different users, or different slots, are
mutually independent.  We let $\xv[t]$ denote the vector signal
transmitted within slot $t$.  The received signal of the $n$-\/th user
at time $t$ is given by
\[
y_n[t]= z_n[t]+v_n[t]
\]
where $v_n[t]$ is $\Cc\Nc(0,1)$, i.i.d. in $n$ and $t$, and where
\[
z_n[t] = \hv^\herm_n[t]\xv[t]\,
\]
with $\hv_n[t]$ denoting the vector channel between the transmitter
and user $n$ in slot $t$.
In all the succeeding sections, we
assume that a subset of $K$ out of $L$ users are scheduled for
simultaneous transmission with $K=M$.

\subsection{MAT Scheme with M=K=2}
Next we give a brief description of the two-user MAT scheme
\cite{maddah2010completely}. The scheme requires $M=2$ BS antennas and
serves $K=2$ users with a two-round scheme. The first round uses two slots, each
used by the BS to transmit a message intended for a single user.
The second round uses a single slot and contains a
message simultaneously useful to both users.  In particular, in round-1 in slot
$t=j$ with $j=1,2$, the BS sends a $2\times 1$ vector symbol
$\bfx{x}_j$ intended for user j (i.e., $\mathbf{x}[j] = \mathbf{x}_j$
for $j=1,2$).  As a result, user $n$ with $n=1,2$ receives the
following observations within slots with $j=1,2$:
\begin{equation}
y_n[j] = \mathbf{h}_n[j]^\herm\mathbf{x}_j + v_n[j],\ \ \ \  j=1,\, 2
\label{round-1-obs}
\end{equation}
After round-1 the $n$-th user has one scalar observation of its
intended
message. It also has one scalar observation of the message intended for the other user, for which it
is simply an eavesdropper.  The second round transmission occurs
within a third slot labeled slot $t=3$, and
consists of a message simultaneously useful to both users. In
particular, the BS forms a new scalar symbol (stream) that equals the
sum of the two-users scalar eavesdropped observations:
\begin{equation}
x_{1,2} = \mathbf{h}_1[2]^\herm\mathbf{x}_2 +
\mathbf{h}_2[1]^\herm\mathbf{x}_1
\end{equation}
The BS transmits the scalar over one linear dimension, e.g.~over antenna 1.  User $n$ obtains
the following observation:
\begin{equation}
y_n[3] = \alpha h_n[3] x_{1,2} + v_n[3]
\label{Slot-3-obs}
\end{equation}
and where $h_n[3]$ denotes the scalar channel between transmit antenna
1 and user n in slot 3.  The parameter $\alpha = \frac{1}{\sqrt{2}}$
ensures that the average power constraint is satisfied.

Using the observations from the three slots, and after
canceling out interference, each user sees an equivalent $2 \times 2$
channel and can thus decode its own message. For
example, user 1 obtains
\begin{eqnarray}
\left[
\begin{array}{c}
y_1[1]\\
\!\!\!y_1[3] - \alpha h_1[3]y_1[2]\!\!
\end{array}
\right] = \left[\begin{array}{c} \mathbf{h}_1[1]^\herm\\
\!\!\alpha h_1[3]\mathbf{h}_2[1]^\herm
\end{array}\!\!\right]\mathbf{x}_1 + \left[
\begin{array}{c}
\!v_1[1]\!\\
\!\tilde{v}_1[2]\!
\end{array}
\right]
\end{eqnarray}
whereby $ \tilde{v}_1[2]= v_1[3] - \alpha h_1[3]v_1[2]$.  Thus, each
user is able to decode two  symbols over 3 slots, yielding
DoF$=\frac{4}{3}$.

We note, that in order to enable the round-2 slot (third) transmission
the transmitter needs to have available the round-1 eavesdropper
channels, $\mathbf{h}_2[1]$ and $\mathbf{h}_1[2]$. Therefore it is assumed
that the third slot (associated with coherence block 3) occurs
sufficiently later that the first two slots, to allow for users 1 and
2 to feed their CSI to the transmitter. Furthermore, and implicit in the DoF
calculations, is the assumption that the intended user of message $i$ has also
available to it the CSI seem by the eavesdropper during the round 1 transmission
of message $i$. That is, user 1 has available $\mathbf{h}_2[1]$, i.e.,
  user 2's channel during slot 1, and user 2 has available
  $\mathbf{h}_1[2]$, i.e., user 1's channel during slot 2. Hence, to
    enable this MAT scheme, each eavesdropper channel needs to be
    communicated by the appropriate eavesdropper {\em both} to the BS (in
    order to enable the MAT scheme transmissions) and to the intended
    receiver of the eavesdropped transmission.

\subsection{Brief Description of 3-User MAT Schemes}

The 3-user MAT schemes \cite{maddah2010completely} build upon the
principles of the 2-user MAT scheme. They require at least three
antennas at the BS to serve 3 users in a multi-round transmission, and can be operated with
 either 2 or 3 rounds.

The 2-round 3-user MAT scheme uses 3 slots in the first round and 3 slots in the second round.
In the first round, slot
$j$ for $j=1,\,2,\,3$ is used to transmit a 3-dimensional symbol
$\mathbf{x}_j$ to user $j$.  After round-1, each user has one scalar
observation of its own message and two scalar eavesdropped
observations. In the second round, 3 scalar messages of the form $x_{i,j}$ are
formed at the BS: $x_{i,j}$ is the sum of the (eavesdropped)
observations collected by users $i$ and $j$ from the round-1 slot
transmission of each other's message. The BS then uses a slot to
transmit each such degree-2 message. Using its round-2 observations,
each user can then strip out the two round-1 eavesdropper observations
of its own message. This together with the user's own round-1
observation of its message allow the user to decode its message. This
scheme yields 9 symbols over 6 slots and thus a DoF$=1.5$.

The 3-round scheme uses 6 slots in round-1, 3 slots in round-2, and 2 slots in round-3.
In round-1 the BS uses for each user 2 slots to transmit two 3-dimensional messages to each
user. Thus each user $i$ then has two (round-1) eavesdropped scalar observations
for messages intended for user $j$, $j\neq i$. As a result, the degree-2 messages
formed after round-1 at the BS, of the form $\mathbf{x}_{i,j}$, are
two-dimensional. Each such degree-2 message, i.e., $\mathbf{x}_{i,j}$,
is transmitted once over a round-2 slot (total of 3 round-2 slots).

After round-2, an intended recipient of message $\xv_{i,j}$, e.g., user $i$, now requires a single round-2
eavesdropper observation to decode this message.   This observation  is made available to user $i$ in
round 3. In particular, the three round-2 scalar eavesdropper observations
(one per user or per round-2 slot) are used to generate a single
3-dimensional degree-3 message. The BS uses two round-3 slots to
transmit two linear combinations of this message (round 3). Based on
its two round-3 observations, each user can decode the 2
scalar elements (out of the 3 in the degree-3 message) that it does not
have. This allows user $j$ to also decode the degree-2 messages
intended for the user and in turn decode its own 6-dimensional
message. This scheme yields 18 symbols over $6+3+2=11$ slots and thus
DoF$=18/11=1.636$.  Note, the three round scheme does have higher DoF than the two round scheme.  However, the difference is small.

\subsection{Brief Description of $K$-User MAT Schemes}

The $K$-user $K$-round MAT scheme \cite{maddah2010completely} uses a
BS with at least $K$ transmit antennas to simultaneously serve $K$
single-antenna users by means of $K$ rounds of transmissions.  The
first round consists of $Q$ slots,
where $Q$ is some properly chosen integer\footnote{The value of $Q$ is
  a multiple of $K!$, i.e., it is such that the number of
  transmissions required in each round for each degree-$r$ message
  intended for each user $r$-tuple is an integer.}.  Round $r$ of the
protocol comprises of $Q/r$ slots: based on CSIT from round $r-1$, the
BS generates $Q/r$ degree-$r$ messages and transmits them over $Q/r$
slots.  Note that the CSIT required from a round-$r$ message, i.e., a
message simultaneously useful to $r$ users, is the CSI of all $K-r$
eavesdropping users during the transmission slot of that message. This
is then used at the BS to regenerate the eavesdropper observations
(without the noise) and in turn, degree-$(r+1)$ messages for round
$r+1$.

Although the $K$-round scheme results in the maximum DoFs, schemes
with $R$: $2\le R<K$ rounds are also attractive (as seen in the $K=3$ examples). The first
$R-1$ rounds of an $R$-round scheme are identical to those of the
$K$-round scheme. The last ($R$-th) round in this case, however,
consists of $Q(K+1-R)/R$ transmissions of scalar degree-$R$ messages.

\section{Achievable Rates with Training}
\label{csi-est-fb}

In this section, we analyze the performance of the MAT scheme taking
into account  aspects of training and feedback. The analysis is
based on immediate extensions of the approach in
\cite{Caire-Jindal-Kobayashi-Ravindran-TIT10}. For simplicity, we focus on the MAT scheme
for $M=K=2$ users.  The case $M>2$ can be handled with
straightforward, albeit tedious, extensions of the $M=2$ case.

The $2$-user MAT scheme  requires the following:
\begin{itemize}
\item[A.] Downlink training (per slot): This allows each user to
  estimate its channel in any given slot.
\item[B.] Channel state feedback: This allows eavesdropper channel CSI of the respective UT
  in any given slot to be made available to the BS and the intended (other)
  receiver.
\item[C.] Data transmission and decoding: This includes: the round-1
  slot transmissions;  generation and transmission of the round-2 messages; and decoding at each user.
\end{itemize}

\subsection{Downlink Training}
In order to enable  channel estimation in round-1 slots, $\beta_1M$ shared pilots
($\beta_1 \geq 1$ symbols per antenna) are transmitted in the
downlink in each slot. UT $k$ for $k \in \{1,2\}$ estimates its slot-$j$ channel
from the observation
\begin{equation}
\mathbf{s}_k[j] = \sqrt{\beta_1P}\mathbf{h}_k [j]+ \mathbf{v}_k[j]
\end{equation}
and where $\mathbf{v}_k \sim \mathcal{CN}(0,N_0\mathbf{I})$. The MMSE
estimate of user $k$'s channel in slot $j$ is given as
\begin{equation}
\tilde{\mathbf{h}}_k[j] =
\EE[\bfx{h}_k[j]\bfx{s}_k^\herm[j]]\EE[\bfx{s}_k[j]\bfx{s}_k^\herm[j]]^{-1}\bfx{s}_k [j]=
\frac{\sqrt{\beta_1P}}{N_0 + \beta_1P}\bfx{s}_k[j]
\end{equation}
Note that $\bfx{h}_k[j]$ can be written in terms of the estimate
$\tilde{\bfx{h}}_k[j]$ and independent white Gaussian noise $\bfx{n}_k[j]$
as \cite{Caire-Jindal-Kobayashi-Ravindran-TIT10}:
\begin{equation}
\bfx{h}_k[j] = \tilde{\mathbf{h}}_k[j] + \bfx{n}_k[j]
\label{htilde-mmse-orth}
\end{equation}
where $\bfx{n}_k[j]$ is Gaussian with covariance:
\begin{equation}
\EE[\bfx{n}_k[j]\bfx{n}_k^\herm[j]] =
\mbox{$\sigma_1^2\bfx{I}$, with $\sigma_1^2 = \frac{1}{1 + \beta_1P/N_0}$}
\end{equation}

\subsection{Channel State Feedback}
To enable the round-2 transmission, each user has to feed back to the
BS its own channel seen during the round-1 slot for which it was an
eavesdropper. This channel needs to also be communicated to the
intended user of message (i.e., the other user).  We use
$\hat{\bfx{H}} = [\hat{\bfx{h}}_1[2],\hat{\bfx{h}}_2[1]] \in \mathbb{C}^{2
  \times 2}$ to denote the imperfect eavesdropper CSI available at the BS
corresponding to the true channel $\bfx{H}= [{\bfx{h}}_1[2],{\bfx{h}}_2[1]]$.


This dual training for CSIT and CSIR can be accomplished in various
ways.  In this paper we assume that it is accomplished by letting
users take turns in time (in a round-robin fashion) to  feed back their
CSI to the BS.  When a particular user is
transmitting, all other users are silent and thus other UTs can also receive
the CSI feedback. This is be best suited to situations when the users are
sufficiently closely located, i.e., the channel between the user
sending the feedback and the user listening is very
strong\footnote{The general problem of efficient CSIT and CSIR
  dissemination is beyond the scope of this paper.}.


We assume analog feedback and  make the simplifying assumption that the feedback channel is
unfaded AWGN, with the same downlink SNR, $P/N_0$, and that the UTs
make use of orthogonal signaling. The number of feedback symbols per
antenna is given by $\beta_f$.

Recall that each UT receives $\bfx{s}_k[j] = \sqrt{\beta_1P}\bfx{h}_k[j] +
\bfx{v}_k[j]$ during the downlink training phase. Then, each UT
transmits a scaled version of $\bfx{s}_k[j]$ during the channel
feedback phase and the resulting observation at the BS is given by
\begin{eqnarray}
\bfx{g}_{{\rm BS},k}[j] &=& \frac{\sqrt{\beta_fP}}{\sqrt{\beta_1P \!+\!
N_0}}\bfx{s}_k [j]+ \tilde{\bfx{w}}_k[j] \nonumber\\
&=& \frac{\sqrt{\beta_f \beta_1 P^2}}{\sqrt{\beta_1P \!+\!
N_0}}\bfx{h}_k[j] + \frac{\sqrt{\beta_f P}}{\sqrt{\beta_1P \!+\!
N_0}}\bfx{v}_k[j] +
\tilde{\bfx{w}}_k[j] \nonumber\\
&=& \frac{\sqrt{\beta_f \beta_1 P^2}}{\sqrt{\beta_1P \!+\!
N_0}}\bfx{h}_k[j] + \bfx{w}_k[j]
\end{eqnarray}
where $\tilde{\bfx{w}}_k$ represents the AWGN noise on the uplink
feedback channel (variance $N_0$) and $\bfx{v}_k$ is the noise
during the downlink training phase. Following the analysis of
\cite{Caire-Jindal-Kobayashi-Ravindran-TIT10}, we can write $\bfx{h}_k$ in terms of
$\hat{\bfx{h}}_k$ as follows:
\begin{equation}
\bfx{h}_k[j] = \hat{\bfx{h}}_k[j] + \bfx{e}_k[j]
\end{equation}
where $\hat{\bfx{h}}_k[j]$ and $\bfx{e}_k[j]$ are mutually independent and
$\bfx{e}_k$ has Gaussian i.i.d. components with zero mean and
variance:
\begin{equation}
\sigma_e^2 = \sigma_w^2/\mbox{$(\sigma_w^2 +
\frac{\beta_f\beta_1P^2}{\beta_1P + N_0})$   }
\end{equation}
with $\sigma_w^2 = N_0(1 + \frac{\beta_fP/N_0}{1 + \beta_1P/N_0})$.

We assume that the feedback channel between users has a different SNR,
given by $P_1/N_0$, which quantifies the strength of the channel (for
example, if the users are close $P_1 \gg P_0$). Proceeding on similar
grounds, the MMSE estimate of the channel vector $\bfx{h}_k[j]$ is
given by
\begin{equation}
\bfx{h}_k[j] = \breve{\bfx{h}}_k[j] + \bfx{f}_k[j]
\end{equation}
where $\breve{\bfx{h}}_k[j]$ and $\bfx{f}_k[j]$ are mutually independent
and $\bfx{f}_k[j]$ has Gaussian i.i.d. components with zero mean and
variance:
\begin{equation}
\sigma_f^2 = \sigma_x^2/ (\sigma_x^2 +\mbox{$
\frac{\beta_f P_1\beta_1P}{\beta_1P + N_0}$})
\end{equation}
with $\sigma_x^2 = N_0(1 + \frac{\beta_fP_1/N_0}{1 +
\beta_1P/N_0})$.


\subsection{Data Transmission}

In the data-transmission portion of each slot, the BS transmits
messages that comprise coded data symbols. Each such message is
received by both receivers. Without loss of generality we focus on
user 1. Using (\ref{round-1-obs}) and (\ref{htilde-mmse-orth}), the
observations of user 1 in slots $j=1,2$ can be expressed as follows:
\begin{equation}
y_1[j]
= \tilde{\bfx{h}}_1[j]^\herm\bfx{x}_j + \bfx{n}_1[j]^\herm \bfx{x}_j +
v_1[j]\ \ j=1,2
\end{equation}
%
The BS then uses its round-1 CSIT to form the scalar message
($\hat{\bfx{h}}_1[2]^\herm\bfx{x}_2 +
\hat{\bfx{h}}_2[1]^\herm\bfx{x}_1$) and transmits it in the third
slot. The resulting user-1 observation (\ref{Slot-3-obs}) can be
expressed as
\begin{eqnarray}
y_1[3]
&=& \alpha \tilde{h}_1[3] \{\hat{\bfx{h}}_1[2]^\herm\bfx{x}_2 +
\hat{\bfx{h}}_2[1]^\herm\bfx{x}_1\} + \nonumber\\
&& \alpha n_1[3] \{\hat{\bfx{h}}_1[2]^\herm\bfx{x}_2 +
\hat{\bfx{h}}_2[1]^\herm\bfx{x}_1\} + v_1[3]
\label{y13-alt-exp}
\end{eqnarray}
where $\alpha$ is a power normalization\footnote{Note here that
  $\alpha > \frac{1}{\sqrt{2}}$ can be used, since the BS has
  estimates of the channels whose elements have power $<$ 1.   However, for sufficiently high SNR
  or large $\beta$, the expected power is close to 1 with $\alpha$ close to
  $\frac{1}{\sqrt{2}}$.}, which ensures that the transmitted symbols
satisfy the average power constraint\footnote{In this phase, the BS
  sends pilot symbols to train the user channels
  $\textbf{h}_1[3]$ using only one transmit antenna.  Thus $h_1[3]$ is a scalar term.}.

User 1 has the estimates $\breve{\bfx{h}}_2[1]$ and
$\tilde{\bfx{h}}_1[2]$ of the true channels $\bfx{h}_2[1]$ and
$\bfx{h}_1[2]$. It therefore needs to compute the MMSE estimates of
$\hat{\bfx{h}}_1[2]$ given $\tilde{\bfx{h}}_1[2]$, and of
$\hat{\bfx{h}}_2[1]$ given $\breve{\bfx{h}}_2[1]$. Applying the
results of MMSE estimation theory, we obtain:
\begin{eqnarray}
\hat{\bfx{h}}_1[2] = \check{\bfx{h}}_1[2] + \bfx{\zeta}_1[2] \nonumber\\
\hat{\bfx{h}}_2[1] = \check{\bfx{h}}_2[1] + \bfx{\zeta}_2[1]
\label{hhat-hcheck-zeta-exp}
\end{eqnarray}
\begin{eqnarray}
\hspace{-0.25in}\mbox{where} \hspace{0.25in} \check{\bfx{h}}_1[2] = \EE[\hat{\bfx{h}}_1[2]|\tilde{\bfx{h}}_1[2]]
\!\!\!&=&\!\!\! \gamma \, \tilde{\bfx{h}}_1[2] \nonumber\\
\check{\bfx{h}}_2[1] = \EE[\hat{\bfx{h}}_2[1]|\breve{\bfx{h}}_2[1]]
\!\!\!&=&\!\!\! \gamma \,\breve{\bfx{h}}_2[1]
\label{hcheck-exp}
\end{eqnarray}
with $\gamma = \frac{\beta_fP}{\beta_fP + N_0}$ and
$\bfx{\zeta}_1[2]$ and $\bfx{\zeta}_2[1]$ have i.i.d. components
with variances given by $\sigma_a^2$ and $\sigma_b^2$ respectively.
\begin{eqnarray}
\sigma_a^2 &\!\!\!\!=\!\!\!\!& \frac{\beta_fP}{\beta_fP +
N_0}\frac{\beta_1P}{\beta_1P
+ N_0}\left(1 - \frac{\beta_fP}{\beta_fP + N_0}\right) \nonumber\\
\sigma_b^2 &\!\!\!\!=\!\!\!\!& \frac{\beta_fP}{\beta_fP \!+\!
N_0}\frac{\beta_1P}{\beta_1P \!+\! N_0}\left(1 -
\frac{\beta_fP}{\beta_fP \!+ \!N_0}\frac{\beta_fP_1}{\beta_fP_1 \!+\!
N_0}\right)\ \ \ \ \
\end{eqnarray}
Using (\ref{hhat-hcheck-zeta-exp}) and (\ref{hcheck-exp}), we can re-express (\ref{y13-alt-exp}) as follows:
\begin{eqnarray}
y_1[3]
&\!\!\!\!=\!\!\!\!& \alpha \tilde{h}_1[3] \{\check{\bfx{h}}_1[2]^\herm\bfx{x}_2 +
\check{\bfx{h}}_2[1]^\herm\bfx{x}_1\} + \nonumber\\
&\!\!\!\!\!\!\!\!& \alpha n_1[3] \{\check{\bfx{h}}_1[2]^\herm\bfx{x}_2 +
\check{\bfx{h}}_2[1]^\herm\bfx{x}_1\} + \nonumber\\
&\!\!\!\!\!\!\!\!& \alpha (\tilde{h}_1[3] + n_1[3])(\bfx{\zeta}_1[2]^\herm\bfx{x}_2 +
\bfx{\zeta}_2[1]^\herm\bfx{x}_1) + v_1[3]\ \ \ \ \
\end{eqnarray}
The effective output for user 1 (after cancelation of the undesired
signal) is then given by
\begin{eqnarray}
\label{decode-eqn-user-1} \left[
\begin{array}{c}
y_1[1]\\
y_1[3] - \alpha \gamma \tilde{h}_1[3]y_1[2]
\end{array}
\right] = \left[\begin{array}{c} \tilde{\mathbf{h}}_1[1]^\herm\\
\alpha \tilde{h}_1[3]\check{\mathbf{h}}_2[1]^\herm
\end{array}\right]\mathbf{x}_1 + \nonumber\\
\bfx{Bx}_2 + \bfx{I_1}\bfx{x}_1 + \bfx{I}_2\bfx{x}_2 + \left[
\begin{array}{c}
v_1[1]\\
v_1[3] - \alpha \tilde{h}_1[3]v_1[2]
\end{array}
\right]
\end{eqnarray}
where
\begin{eqnarray}
\bfx{B} &\!\!\!\!=\!\!\!\!& \left[\begin{array}{c} \mathbf{0}\\
\alpha \tilde{h}_1[3](\check{\mathbf{h}}_1[2]^\herm - \gamma
\tilde{\mathbf{h}}_1[2]^\herm)
\end{array}\right]\\
\bfx{I}_1 &\!\!\!\!=\!\!\!\!& \left[\begin{array}{c} \bfx{n}_1[1]^\herm\\
\alpha n_1[3](\check{\mathbf{h}}_2[1]^\herm + \bfx{\zeta}_2[1]^\herm) +
\alpha \tilde{h}_1[3]\bfx{\zeta}_2[1]^\herm
\end{array}\right]\\
\bfx{I}_2 &\!\!\!\!=\!\!\!\!& \alpha\!\!\left[\!\!\!\begin{array}{c}
  \mathbf{0}\\ n_1[3](\check{\mathbf{h}}_1[2]^\herm \!\!+\!
  \bfx{\zeta}_1[2]^\herm) \!+\!  \tilde{h}_1[3](\bfx{\zeta}_1[2]^\herm
  \!\!-\! \gamma \bfx{n}_1[2]^\herm)
\end{array}\!\!\!\right]\ \ \ \
\end{eqnarray}

\subsection{Achievable Rate Bounds}
\label{rate-achievable}

Bounds on the achievable rates that can be obtained with the MAT
scheme on the basis of downlink training and analog feedback can be
readily derived.  We next derive a lower bound on the mutual
information of user 1, denoted by $\mathcal{R}_1$, assuming Gaussian
inputs, i.e., $x_k \sim \mathcal{CN}(0,P/M)$.
From (\ref{decode-eqn-user-1}) we have
\begin{equation}
y = \bfx{A}\bfx{x}_1 + \bfx{Bx}_2 + \bfx{I_1}\bfx{x}_1 +
\bfx{I}_2\bfx{x}_2 + \bfx{v}
\end{equation}
The achievable rate with Gaussian inputs and CSI training and
feedback is lower bounded by
\begin{eqnarray}
\mathcal{R}_1 &\geq& \frac{2}{3}\EE\left[\log  \left|\bfx{N}_{\rm
MAT} +
(\bfx{A}\bfx{A}^\herm + \bfx{B}\bfx{B}^\herm + \bfx{I}_A + \bfx{I}_B)P/M\right| \right.\nonumber\\
&& \left.- \log \left| \bfx{N}_{\rm MAT} + (\bfx{B}\bfx{B}^\herm +
\bfx{I}_A + \bfx{I}_B)P/M\right| \right]
\end{eqnarray}
where
\begin{eqnarray}
\bfx{A} \!=\!\left[\!\!\begin{array}{c} \tilde{\mathbf{h}}_1[1]^\herm\\
\alpha \tilde{h}_1[3]\check{\mathbf{h}}_2[1]^\herm
\end{array}\!\!\right] \!\!\!&&\!\!\!\!\!
\bfx{N}_{\rm MAT}\! \!=\!\left[\!\!\begin{array}{cc} N_0 &\!\! 0\\
0 & \!\!\! N_0(1 + |\alpha \gamma \tilde{h}_1[3]|^2)
\end{array}\!\!\right] \nonumber
\end{eqnarray}
\begin{eqnarray}
\bfx{I}_A
&\!\!\!\!\!=\!\!\!\!\!& \left[\begin{array}{cc} \!M\sigma_1^2\! & 0\\
\!0 &\! \alpha^2(\sigma_1^2(M\sigma_b^2 + ||\check{\bfx{h}}_2[1]||^2) +
M|\tilde{h}_1[3]|^2\sigma_b^2))
\end{array}\!\right]\nonumber\\
\bfx{I}_B
&\!\!\!\!\!=\!\!\!\!\!& \left[\begin{array}{cc}\!\! 0\!\! & 0\\
\!\!0\!\! & \alpha^2(\sigma_1^2(M\sigma_a^2 \!+\! \|\check{\bfx{h}}_1[2]\|^2)\! +\!
M|\tilde{h}_1[3]|^2(\sigma_a^2 \!+\! \gamma^2\!\sigma_1^2))\!
\end{array}\!\!\!\right] \nonumber
\end{eqnarray}
The proof of the above result follows from
\cite{Caire-Jindal-Kobayashi-Ravindran-TIT10}.  Bounds on rates for user-2 follow and are the same as user-1.

\section{Scheduling}
\label{scheduling}

 The scheduling setting we consider involves an $M$-antenna
 transmitter and $L$ single antenna users.  We let $\xv_m(i)$ denote
 the $i$-\/th (coded) message intended for user $m$.  We also let
 $t_m(i)$ denote the index of the (first-round) slot over which
 message $\xv_m(i)$ is transmitted, i.e., $\xv[t_m(i)]=\xv_m(i)$.


We consider scheduling algorithms in the family derived via stochastic
optimization using the Liapunov drift technique
\cite{Shirani-Mehr-Caire-Neely-TCOMM10}, according to which, at each
scheduling slot, $t$, the scheduler updates ``weights'' for each user
and solves a max-weight sum-rate maximization problem. The weights can
be interpreted as the backlog of some appropriately designed ``virtual
queues,'' that play the role of stochastic versions of Lagrangian
multipliers in the associated network utility function maximization
problem.  The scheduling decision in slot $t$ exploits knowledge of
the transmitter-user channels, over all $\tau$: $\tau<t$.  For
simplicity, we assume that, for each past transmission that CSIT is
available for scheduling from all $L$ users.
Also, we focus on the case where all users have the same SNR and the
scheduling criterion is the expected sum user-rate\footnote{The
  general unequal SNR case with a general system-wide utility metric
  can be similarly captured with appropriate
  extensions\cite{Shirani-Mehr-Caire-Neely-TCOMM10}.}.

In order to appreciate the potential challenges and benefits of
scheduling for MU-MIMO with outdated CSI, it is worth contrasting it
against scheduling for conventional MU-MIMO. In conventional MU-MIMO,
CSIT is collected about the channels between the transmitter and
multiple users. The scheduler at the transmitter uses this CSIT to
select a subset of users for MU-MIMO transmission along with a precoder.
The  assumption with scheduling conventional MU-MIMO is that the
channels based on which CSIT is obtained and the channels over which
the MU-MIMO transmission takes place are sufficiently correlated
(they differ by an error with a sufficiently small variance)
\cite{Caire-Jindal-Kobayashi-Ravindran-TIT10}.

Much like with conventional MU-MIMO, CSI from multiple users can be
exploited to schedule joint MU-MIMO transmissions with outdated CSI
 to optimize some system utility metric\footnote{Scheduling requires the CSI of a UT in slots for which it receives its intended message. This is an added requirement over the basic MAT scheme.}. The key
difference here is that these are multi-round schemes, whereby the
MU-MIMO transmissions at a given round are ``joint'' transmissions
of several eavesdropped messages from the previous rounds, and thus
only exploit CSIT from past rounds. Furthermore, as all CSIT available is from past transmission slots only, the exact timing of
the scheduled transmissions does not matter.

We first consider MAT session schedulers, i.e., schedulers that
schedule packets from different sets of users into MAT sessions. We
then consider a class of schedulers that schedule multi-round MU-MIMO
transmissions based on outdated CSI in a more flexible manner.

\subsection{MAT-Session Scheduling}
We first consider the 2-user MAT (MAT-2) session scheduling problem in
detail. We then briefly comment on extensions for the 2-round $K$-user
problem and then the R-round K-user scheduling sessions, with $R\le
K$, $K\le M$ and $K\le L$.

The 2-user MAT-session scheduler schedules pairs of user packets of
the form $(\xv_m(i), \xv_n(j))$ with $m\ne n$ in two-round MAT
sessions. Note that, since the round-1 transmissions involve {\em
  individual} user messages, the pairing decisions need only occur just prior to
the second round transmission.  Pairing involves the sum of the
eavesdropped observations from first-round transmissions.

Given a MAT session between the $i$-\/th packet of user $m$ and the
$j$-\/th packet of user $n$, its round-2 slot is denoted by
$t_{m,n}(i,j)$, and satisfies $t_{m,n}(i,j) > \max \{ t_{m}(i),
t_{n}(j)\}$. The associated transmitted signal is given by
\[
\xv[t_{m,\!n}(i,\!j)] \! = \! \begin{bmatrix}1 \\ 0 \end{bmatrix}
\!\alpha_{m,\!n}(i,\!j)
\!\left( \hv^\herm_n[t_m(i)]\xv_m(i) \! +\!  \hv^\herm_m[t_n(j)]\xv_n(j)  \right)
\]
whereby the scaling constant $\alpha_{m,n}(i,j)$ is chosen so as to
ensure constant power transmission, i.e.,
\[
\alpha_{m,n}(i,j) = \mbox{$\frac{\sqrt{2}}{\sqrt{\|\hv^\herm_n[t_m(i)]\|^2+ \|\hv^\herm_m[t_n(j)]\|^2 }}$}.
\]

For convenience, we focus on a fixed-buffer size scheduler. In
particular, we assume that at each scheduling instance (i.e., each
time a round-2 transmission is to be scheduled) the scheduler has
available CSI from all $L$ users on $LN$ round-1 slots, and exactly
$N$ of these slots carried messages for a given user.  Once a round-2
transmission is scheduled between some packet $i$ of some user $m$ and
some packet $j$ of some user $n$, this transmission is also
accompanied by two new round-1 transmissions of fresh packets to users $m$
and $n$.

The optimal scheduling algorithm in this case is then straightforward
to derive.  At any given scheduling instance, the scheduler has CSIT
for packets $\xv_m(i)$ for $1\le i\le N$ and $1\le m\le L$ (without
loss of generality the packet indices of each user are indexed from 1
to $N$).  The scheduling problem reduces to the following optimization
\cite{Shirani-Mehr-Caire-Neely-TCOMM10}
\begin{equation}
(m^*\!,i^*\!,n^*\!,j^*)  = \!\!\! \!\! \!\!\!\! \!\!\argmax_{\stackrel{(m,i,n,j):}{\tiny 1\le m<n\le L,\, 1 \le i,j \le N}}\!\!\!\!\!\!\!\!\!\!
Q_m \Delta\bar{R}_{m,i}(n,j) + Q_n \Delta\bar{R}_{n,j}(m,i)
\label{Sched-2user-MAT}
\end{equation}
where $Q_m$ denotes the optimization weight\footnote{The weight,
  $Q_m$, of user $m$ at a given scheduling slot, $t$, is provided to
  the scheduler and is simply the output of the virtual-queue process
  of user $m$ at time $t$ \cite{Shirani-Mehr-Caire-Neely-TCOMM10}.} of
user $m$, and where $\Delta\bar{R}_{m,i}(n,j)$ is the expected
mutual-information increase to user $m$ by performing a round-2
transmission that completes an existing (in progress) MAT-2 session between $\xv_m(i)$ and
$\xv_n(j)$.  This expected increase  is given by
\[
\Delta\bar{R}_{m,i}(n,j) = \bar{R}_{m,i}(n,j) - R_{m,i} + \bar{R}_{m}
\]
where
\[
\bar{R}_{m,i}(n,j)=\EE_f\log(\det [ I +
\bfx{K}_z^{-1}\bfx{H}_{m,n}\bfx{H}^\herm_{m,n} P/2 ])
\]
is the expected mutual information
provided by the MAT session to user $m$ after performing interference
alignment,
\begin{eqnarray}
\label{channel-sch}
\!\!\!\!\!\!\mbox{with $\bfx{H}_{m,n}$} &\!\!\!\!=\!\!\!\!& \left[\!\!\begin{array}{c} \bfx{h}_m[t_m(i)]\\
\frac{\sqrt{2}f}{\sqrt{\|\hv^\herm_n[t_m(i)]\|^2+
\|\hv^\herm_m[t_n(j)]\|^2 }} \bfx{h}_n[t_m(i)]
\end{array}\!\!\right]\nonumber\\
\bfx{K}_z^{-1} &\!\!\!\!=\!\!\!\!& \left[\!\!\begin{array}{cc} N_0 & \! 0\\
0 & \! N_0(1 \!+ \!\frac{2|f|^2}{\|\hv^\herm_n[t_m(i)]\|^2+
\|\hv^\herm_m[t_n(j)]\|^2 })
\end{array}\!\!\!\! \right]
\end{eqnarray}
The quantity
\[
R_{m,i} = \log\left(1+\frac{P}{2N_o}\|\hv_m[t_m(i)]\|^2\right)
\]
is the mutual information from the round-1
transmission of packet $i$, and $\bar{R}_{m}$ is the expected mutual information from a round-1
transmission of a new packet for user $m$.





The above scheduling approach can be generalized to involve $K$-user
$R$-round MAT sessions. However, the scheduling benefits are very
limited due to the restrictive eavesdropper nature of the
MAT-session. To see this consider a 3-user 2-round MAT-session
scheduling scheme. In such a scheme, 3 dimensional messages are transmitted
from 3 antennas to 3 users using 3 round-1 slots and 3 round-2 slots.  The
scheduler in this case would choose, for round-2, three-user MAT sessions between
packets $i_1$, $i_2$, $i_3$, of users $m_1$, $m_2$ and $m_3$,
respectively, based on eavesdropper CSIT from the round-1 transmission
of these packets. In particular user $m_k$ gets three looks at its packet,
one through its own channel and two more through the two eavesdropper
channels (all at the same time). The set of these three channels must
constitute a ``good'' $3 \times 3$ channel (in the sense that the
expected rate of user $k$ after the round-2 transmissions has to be
sufficiently high).  Furthermore, this has to simultaneously happen for all 3
users. As a result, the number of scheduling options required to get
simultaneously good rates to all users grows exponentially fast with
the number of users.

Another limitation of MAT-session based scheduling is that the MAT
session is completely determined by the completion of the second
round, regardless of the total number of rounds. Hence, when
scheduling MAT sessions with more than 2 rounds, once the second round
is completed the rest of the session has been fully determined and no
further scheduling benefits are to be expected.

\subsection{Eavesdropper-Based Packet-Centric Scheduling}

In this section we consider a different approach for scheduling
MU-MIMO transmissions with outdated CSI. It is based on enabling
packet-centric (rather than MAT-session based) interference alignment
for efficient MU-MIMO transmission.  This scheme exploits the same
principles as the MAT scheme and achieves the same DoFs as the MAT
scheme.  In particular, consider $R$-round $K$-user protocols with a
packet-centric scheme.  The scheme has the following properties:
\begin{itemize}
\item Much like the $R$-round $K$-user MAT scheme, for each round $r$
  with $1\le r\le R-1$, and for each degree-$r$ message (i.e., a
  message simultaneously useful to $r$ user terminals) that is
  transmitted, a set of $K-r$ eavesdropper observations are
  communicated to each of the $r$ intended receivers, by means of
  ``network-coded'' IA-enabling transmissions in the following rounds;
\item unlike the MAT scheme, however, the $K-r$ eavesdropper
  observations are not preselected based on the MAT session; rather
  they are chosen based on the channel quality of the eavesdropper
  channels.
\end{itemize}

\begin{figure}
\centerline{\hbox{\psfig{figure=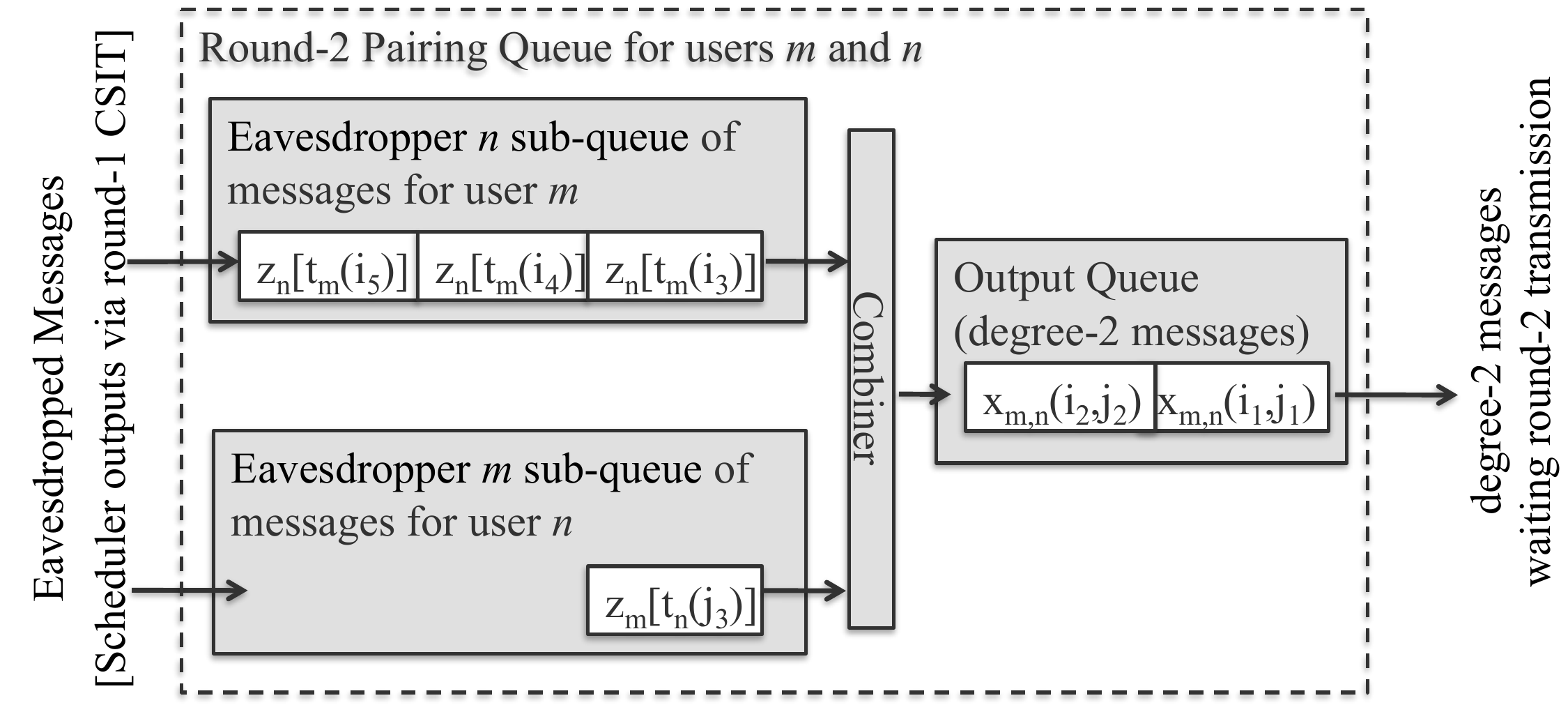, width=3.5in}}}
\caption{Sub-message pairing into degree-2 messages for user pair $(m,n)$.}\label{pairing-queue-2}
\vspace{-0.25in}
\end{figure}
\begin{figure}
\includegraphics[width=3.5in,,bb = 0 210 600 600]{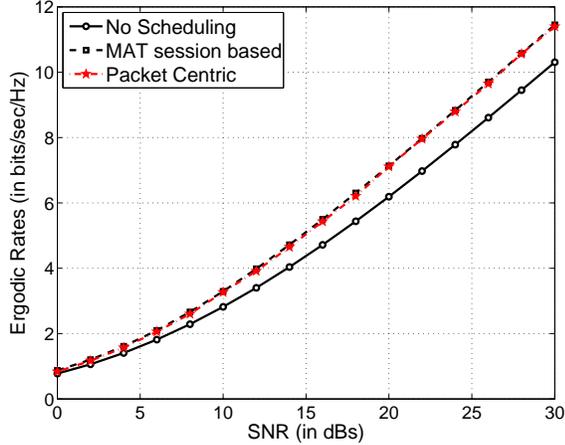}
\caption{MAT-session based vs. packet-centric scheduling: the 2 user case.}\label{pc-vs-session}
\end{figure}

To illustrate the difference between the two schemes consider first
the problem of scheduling round-2 transmissions for a 2-round $K$-user
MU-MIMO packet-centric scheme. This scheme relies on the use of a set
of $(m,n)$ user-terminal pairing queues of the form shown in
Fig~\ref{pairing-queue-2}, which generate degree-2 messages for
round-2 transmissions. The main principles behind packet-centric
eavesdropper-based scheduling can be summarized as follows:
\begin{enumerate}
\item Each round-1 slot involves transmitting a $K$ dimensional
  message intended for one of the $L$ users.
\item For each round-1 transmission intended for a given user,
  say user $m$, the base-station chooses $K-1$ out of the $L-1$
  eavesdroppers for round-2 transmissions (based on round-1
  eavesdropper CSIT).
\item For each such eavesdropper, e.g. eavesdropper $n$, the
  base-station places the eavesdropped observation of user $n$ in the
  corresponding $(m,n)$ queue, in the queue input associated with
  eavesdropper $n$.
\item Degree-2 messages for the user pair $(m,n)$ are formed by
  combining sub-messages from the  queues of eavesdroppers $n$  and
  $m$ within the $(m,n)$ queue. These messages
  then simply wait for (round-2) transmission.
\end{enumerate}
It is interesting to contrast eavesdropper scheduling with MAT-sessions scheduling in the case $K=2$. In this case, and given CSIT (from all users) from the round-1 transmission of message $i$ for user $m$,  the eavesdropper scheduler in step 2) above selects one eavesdropper out of all the users via
\begin{equation}
\label{Sched-2user-pack} \tilde{n}^*(m,i) = {{\rm arg}\max}_{{\tiny
n : 1\le n \le L,\ n \neq }} \bar{R}_{m,i}(n)
\end{equation}
where $\bar{R}_{m,i}(n)$ is a heuristic objective function obtained by
replacing $\|\hv^\herm_m[t_n(j)]\|$ with $\|\hv^\herm_n[t_m(i)]\|$ in
(\ref{channel-sch})\footnote{Although this is a heuristic
  approximation, in principle the objective could be validated by
  proper matching of eavesdropper observations at the combiner of the
  $(m,n)$ queue, such that eavesdropper channels of roughly equal
  norms are combined to generate degree-two messages. As
  Fig.~\ref{pc-vs-session} suggests, however, such careful combining
  is not necessary.}:
\[
\bar{R}_{m,i}(n)= \log(\det [ I +
\bfx{\tilde{H}}_{m,n}\bfx{\tilde{H}}^\herm_{m,n} P/2 ])
\]
\begin{eqnarray}
\mbox{with $~~~~\bfx{\tilde{H}}_{m,n}$} &\!\!\!\!=\!\!\!\!& \left[\begin{array}{c} \bfx{h}_m[t_m(i)]\\
\frac{1}{\sqrt{1 + \|\hv^\herm_n[t_m(i)]\|^2 }} \bfx{h}_n[t_m(i)]
\end{array}\right]
\end{eqnarray}

Fig.~\ref{pc-vs-session} shows a performance comparison between the
heuristic packet centric scheduler (\ref{Sched-2user-pack}) and
the MAT-based one (\ref{Sched-2user-MAT}), assuming $L=20$. As the
figure illustrates, both schedulers yield nearly identical
performance. Heuristic approximations of the form
(\ref{Sched-2user-pack}) can be readily used for implementing packet
centric schedulers with $K$-user schemes. where $K>2$.



\begin{table}
\centering \caption{Sample 3-user 2-round MAT-Session Scheduling between packets 6, 13, and 27 of users 1, 2, and 3, respectively}
\vspace{-0.125in}
\begin{tabular}{||c||c|c|c||}
    \hline Messages for: & $\xv_1(6)$ & $\xv_2(13)$ &
    $\xv_3(27)$\\ \hline\hline Round 1 & $\xv_1(6)$ & $\xv_2(13)$ &
    $\xv_3(27)$\\ \hline Round 2
    &\!\begin{tabular}{c}
\!$\xv_{1,2}(6,13)$\!\\\!$\xv_{1,3}(6,27)$\!\end{tabular}\!&\!\begin{tabular}{c}
\!$\xv_{1,2}(6,13)$\!\\\!$\xv_{2,3}(13,27)$\! \end{tabular}\!&\!\begin{tabular}{c}\!$\xv_{1,3}(6,27)$\!\\\!$\xv_{2,3}(13,27)$\!\end{tabular}\!\\ \hline
\end{tabular}
\label{MAT-session-message-table}
\vspace{0.15in}
\centering \caption{Sample 3-user 2-round Packet-Centric Scheduling
  for packets 6, 13, and 27 of users 1, 2, and 3, respectively}
\vspace{-0.125in}
\begin{tabular}{||c||c|c|c||}
    \hline Messages for: & $\xv_1(6)$ & $\xv_2(13)$ &
    $\xv_3(27)$\\ \hline\hline Round 1 & $\xv_1(6)$ & $\xv_2(13)$ &
    $\xv_3(27)$\\ \hline Round 2
    &\!\begin{tabular}{c}
\!$\xv_{1,2}(6,13)$\!\\\!$\xv_{1,3}(6,27)$\!\end{tabular}\!&\!\begin{tabular}{c}
\!$\xv_{1,2}(6,13)$\!\\\!$\xv_{2,4}(13,9)$\! \end{tabular}\!&\!\begin{tabular}{c}\!$\xv_{1,3}(6,27)$\!\\\!$\xv_{3,5}(27,4)$\!\end{tabular}\!\\ \hline
\end{tabular}
\label{pack-session-message-table}
\end{table}
Note that the DoFs of 2-round $K$-user packet-centric sessions are the
same as the DoFs of the associated 2-round $K$-user MAT session.
However, there is significantly more flexibility in scheduling
eavesdroppers.  Tables~\ref{MAT-session-message-table} and
\ref{pack-session-message-table} provide examples of
 scheduling in a 2-round 3-user
MAT session and of scheduling in a 2-round 3-user packet-centric
approach.   Both involve packet 6 of user 1, packet 13 of user 2, and
packet 27 of user 3. The $i$-th column in each table shows all the
transmitted messages associated with the packet of user $i$. As Table~\ref{MAT-session-message-table} shows, all
transmissions in the MAT-session based scheme are determined by the
scheduled MAT session involving the three user-packets. In contrast and as Table
\ref{pack-session-message-table} shows, in the packet-centric scheme,
the packets of user 2 (and 3) are no longer restricted to be included
in transmissions involving packets of user 1 and 3 (1 and 2). Rather,
the eavesdroppers in each case are chosen independently, and it is up
to the pairing queue to group them into degree-two messages.

\begin{figure}
\centerline{\hbox{\psfig{figure=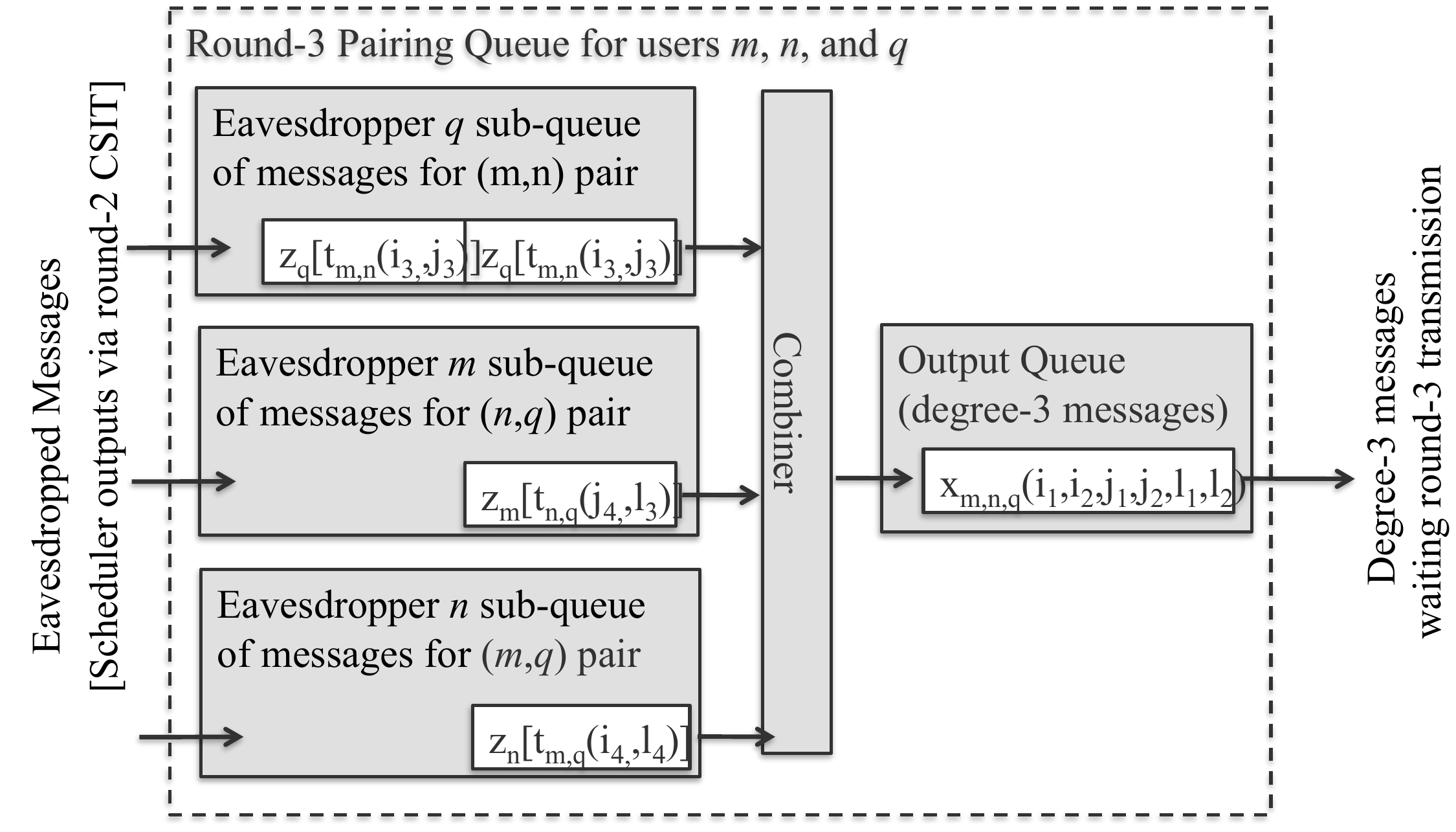, width=3.5in}}}
\caption{Sub-message  combining into degree-3 messages for
  users $(m,n,q)$.}\label{pairing-queue-3}
\vspace{-0.25in}
\end{figure}

The preceding two-round schemes can readily extended to develop
$R$-round $K$-user packet-centric schemes. As an example, consider the
3-round 3-user scheme. In this case, round-2 scheduling uses pairing
queues of the form of Fig.~\ref{pairing-queue-2} and works as already
described. Round-3 scheduling amounts to scheduling eavesdroppers for
degree-2 messages, i.e., messages simultaneously useful to 2
users. Given eavesdropper CSIT from round-2 transmissions intended for
a particular pair of users $(m,n)$, an eavesdropper is selected, e.g.~user $q$, out of all $L-2$ eavesdroppers.  This eavesdropper's message
enters a round-3 pairing queue where it is used to create degree-3
messages for transmission. In particular, it is an input to the user
$q$ eavesdropper queue of the $(m,n,q)$ message queue, shown in
Fig.~\ref{pairing-queue-3}.
As shown in the figure, degree-3 messages (messages simultaneously useful to a triplet of users $(m,n,q)$)
are constructed by combining three eavesdropped observations of degree-two messages, one
for each user eavesdropping on a message intended for the pair of
remaining users.

\section{Simulation Results}
\label{simulation}

In this section we provide a brief performance evaluation of the
MU-MIMO schemes based on outdated CSI.  We first provide a comparison
between the MAT scheme and a conventional MU-MIMO scheme employing
LZFB in the context of training and feedback over time-varying
channels. We assume a block fading channel model that evolves
according to a Gauss Markov process.  The effect of the delay between
the time $t_{pl}$ of downlink pilots (slots in which CSIT is estimated)
and
the time $t_{pl}+t_{\delta}$ of MU-MIMO transmission  is captured by the
magnitude of the expected correlation between
channels from such a pair of time slots.  This coefficient
 is defined by \[\rho(t_{\delta})=\left|\EE_{t}\left[\hv_{k}^{\herm}[t]\hv_{k}[t+t_{\delta}]\right] \right|
 /\, \EE_t\left[ \| \hv_{k}[t]\|^2\right] \]
A value $\rho=1$ means that
the channels are perfectly correlated (co-linear), which happens if the CSI
acquisition and data transmission occur in the same coherence block,
whereas $\rho<1$ indicates that the channels changed between slots.

\begin{figure}
\includegraphics[width=3.5in,bb = 0 210 600
  600]{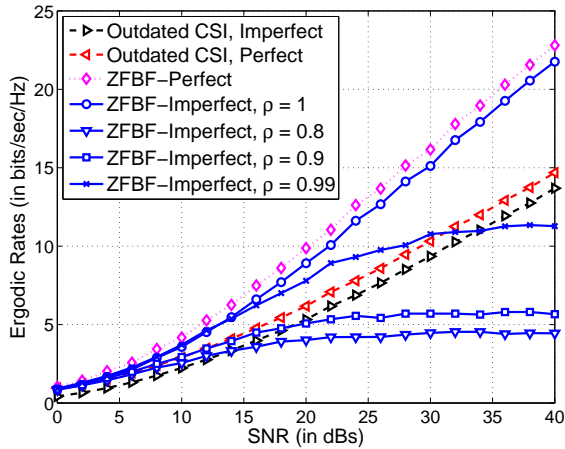} \caption{Comparison between MAT and conventional LZFB for
  a system with 2 antennas at the BS serving two single-antenna
  users.}\label{fig0}
\vspace{-0.25in}
\end{figure}
Fig.~\ref{fig0} shows a comparison between the conventional LZFB and
the MAT scheme for the case of $M = K = 2$ in the presence of
training, and assuming $\beta_1 = \beta_f = 2$ and $P = P_1$.  When
$\rho=1$ LZFB achieves 2 DoFs, as expected
\cite{Caire-Jindal-Kobayashi-Ravindran-TIT10}. There is a constant
rate gap between the perfect CSI case and the case of imperfect CSI based on training, as justified in \cite{Caire-Jindal-Kobayashi-Ravindran-TIT10}.
In contrast for all cases with $\rho<1$, even with a very high correlation of $\rho=0.99$,
the achievable rates eventually saturate as SNR increases
\cite{Caire-Jindal-Kobayashi-Ravindran-TIT10}.  Furthermore, decreasing $\rho$
below $0.99$  results in significantly lower saturation rates.

This rate saturation is not seen with the MAT schemes, which achieve DoF=$\frac{4}{3}$
independent of $\rho$.  Using downlink training
and CSI feedback degrades the achievable rate, however it does so by a constant gap
regardless of the value $\rho$ and similar to what was observed in
\cite{Caire-Jindal-Kobayashi-Ravindran-TIT10} for LZFB when $\rho=1$.


\begin{figure}
\includegraphics[width=3.5in,bb = 0 210 600 600]{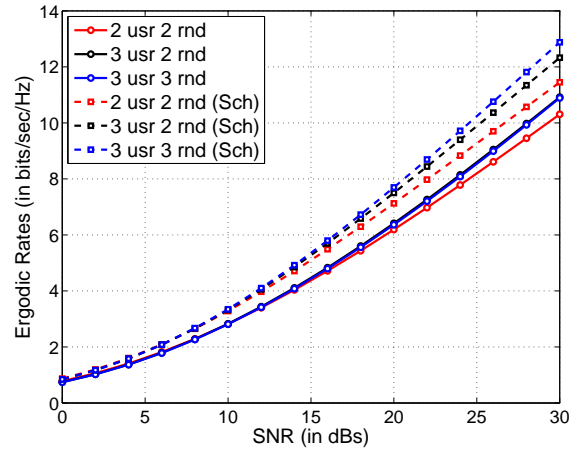}
\caption{Performance of packet-centric scheduling in the case of
  two-user and three-user packet-centric MU-MIMO schemes based on
  outdated CSI.}\label{fig7}
\vspace{-0.125in}
\end{figure}
Fig.~\ref{fig7} shows some of the benefits of packet-centric
scheduling as a function of the number of users served by the MU-MIMO scheme and
the number of rounds used for transmission. In particular, the figure
shows a performance comparison of the ``packet-centric'' based
scheduler for $K = 3$ users with two and three rounds of transmission,
as well as the scheduler's performance for the $K=2$ user case. Also
shown in the figure is the performance of the associated MAT-session
based schemes without scheduling. As the figure suggests, the packet
centric scheduler achieves the DoFs promised by the associated MAT
scheme. In addition, packet-centric scheduling offers more flexibility
when scheduling users, enabling performance benefits to be realized
with a 3-round 3-user scheme at lower SNRs.


\section{Conclusion}
\label{conclusion}

In this paper we considered training and scheduling aspects of
multi-round MU-MIMO schemes that rely on the use of outdated channel
state information (CSI) \cite{maddah2010completely}.
Such schemes are of practical interest as they enable one to rethink many operational aspects of deploying MU-MIMO in dynamic channel conditions, conditions which can inherently limit conventional MU-MIMO approaches.
As shown in the
paper, under proper training, the degrees of freedom promised by these
schemes can be realized even with fully outdated CSI. We also proposed
a novel scheduling algorithm that improves the performance of the
original MAT scheme \cite{maddah2010completely}. It is based on a
variant multi-user MIMO scheme which maintains the MAT scheme DoFs but
provides more scheduling flexibility. As our results suggest, an
appropriately designed MU-MIMO scheme based on multi-round
transmissions and outdated CSI can be a promising technology for
certain practical applications.

\bibliographystyle{IEEEtran}
\bibliography{references}

\end{document}